\pdfoutput=1

\documentclass[11pt]{article}
\usepackage{graphicx}
\graphicspath{ {./images/} }
\usepackage{acl}

\usepackage{times}
\usepackage{latexsym}

\usepackage[T1]{fontenc}

\usepackage[utf8]{inputenc}

\usepackage{microtype}
\usepackage{longtable}
\usepackage{lipsum}   
\usepackage{setspace} 

\usepackage{etoolbox}
\AtBeginEnvironment{quote}{\par\singlespacing\small}

\usepackage[normalem]{ulem} 

\title{Snow Mountain: Dataset of Audio Recordings \\ of The Bible in Low Resource Languages }

\author{Kavitha Raju$^*$, Anjaly V$^*$, Ryan Lish, Joel Mathew \\
\texttt{\{kavitha.raju,anjaly.v,joel\}@bridgeconn.com} \\ \texttt{rslish@cobaltspeech.com}}

\begin{document}
\maketitle

\def\thefootnote{*}\footnotetext{Equal contribution.}\def\thefootnote{\arabic{footnote}}

\begin{abstract}
Automatic Speech Recognition (ASR) has increasing utility in the modern world. There are a many ASR models available for languages with large amounts of training data like English. However, low-resource languages are poorly represented. In response we create and release an open-licensed and formatted dataset\footnote{\url{https://huggingface.co/datasets/bridgeconn/snow-mountain}} of audio recordings of the Bible in low-resource northern Indian languages. We setup multiple experimental splits and train and analyze two competitive ASR models to serve as the baseline for future research using this data. 
\end{abstract}

\section{Introduction}
Automatic Speech Recognition (ASR) is a well-studied problem that is motivated by multiple popular use cases like voice assistants (e.g. Amazon Alexa,  Apple Siri), live audio transcription and as a pre-processing step for conversational machine translation. Training models for ASR, however, are limited by the availability of training data. Thus, fewer ASR models are available for low resource languages which usually do not have a sizable collection of audio recording training data. 

The Bible is a document that is translated into numerous languages including multiple low-resource languages with multiple ongoing translations. There is a parallel effort to produce audio-recordings for the text translations. This is especially important since many of the remaining languages where the Bible is translated are communities with primarily oral learners. In fact, there are translation projects underway that attempt to directly translate the Bible orally without using text. 

Building ASR models for these very low resource languages is an important step to better represent and include these communities in the many technological advancements that are popular among speakers of well represented languages. 

In response to this need, in this paper we describe and share freely-licensed audio recordings of 10 low resource languages of Northern India for The Bible and train two different ASR models using the dataset to serve as a baseline for future research. 

\begin{table*}
\centering
\begin{tabular}{llllllll}
\hline
No. & Language & Bible & Language & Speaker & Speaker & Speaker\\
& & Portion & Code  & ID & Gender & Age\\
\hline
1 & Hindi & OT & hin/hi & Speaker01 & Female & 43-45 yrs\\
2 & Haryanvi & NT & bgc & Speaker02 & Male & 43-45 yrs\\
3 & Bilaspuri & NT & kfs & Speaker03 & Male & 45 -50 yrs\\
4 & Dogri & NT & dgo &  Speaker04 & Male & 29 yrs\\
5 & Bhadrawahi & NT & bhd &  Speaker05 & Male & 30 yrs\\
6 & Gaddi & NT & gbk &  Speaker06 & Male & 30 yrs\\
7 & Kangri & NT & xnr &  Speaker07 & Male & 28-29 yrs\\
8 & Kulvi & NT & kfx &  Speaker08 & Female & 35-40 yrs\\
9 & Mandeali & NT & mjl &  Speaker09 & Female & 20 yrs\\
10 & Kulvi Outer Seraji & NT & kfx-x-OSJ & Speaker10 & Male & 68 yrs\\
11 & Pahari Mahasui & NT & bfz &  Speaker11 & Male & 26-27 yrs\\

\hline
\end{tabular}
\caption{Details of the languages and speakers in the dataset. The listed languages belong to the Indo-Aryan language family. kfx-x-OSJ is a dialect of Kulvi dominant in the region of Outer Seraji in Himachal Pradesh}
\label{tab:languages table}
\end{table*}

\section{Related Work }

Efforts in the development of speech datasets in Tamil, Telugu and Marathi are discussed in  \cite{anumanchipalli2005development} where they collected data from about 560 speakers and trained acoustic models using the Sphinx 2 speech tool kit\cite{CMUSphinx} in the three languages. 

\cite{shrishrimal2012indian} discusses various speech datasets developed in multiple Indian languages for ASR and Text-to-speech models. They collect domain specific data in agriculture, marketing, travel and emergency services. It uses Hindi training data recorded by 30 female speakers with approximately 26 hours of speech recordings. A speech dataset developed for Hindi from news bulletins is mentioned which is 3.5 hours and recorded by 19 speakers (6 Male, 13 Females). Various ongoing projects for creating speech corpora by Linguistic Data Consortium for Indian Languages (LDC-IL) are also underway.

A survey on the efforts made to develop speech corpora in Indian languages is done in \cite{kurian2015review}.

\cite{deka2018speech} present an ongoing effort in creation of speech corpora for under-resourced languages of North-East India, namely, Assamese, Bengali and Nepali.

Our dataset is unique in that it is the first audio recorded corpora available in some of the languages in the dataset while also being substantive in terms of hours of recording when compared with other audio recording datasets for low-resource languages.

\section{Dataset}
The Snow Mountain dataset contains the audio recordings (in .mp3 format) and the corresponding text of The Bible in 11 Indian languages. The recordings were done in a studio setting by native speakers. Each language has a single speaker in the dataset. Most of these languages are geographically concentrated in the Northern part of India around the state of Himachal Pradesh. Being related to Hindi they all use the Devanagari script for transcription. Details of the dataset are shown in tables \ref{tab:languages table} and \ref{tab:datasize table}, including the languages' information \cite{enthnologue}, speaker details and duration of audio. 

The protestant Bible is composed of 66 canonical books which are portions of text of various sizes. These are split into two parts or testaments: the Old Testament (OT) consists of 39 books and the New Testament (NT) consists of 27 books. In the dataset, in the place of using full Bible book names, a 3-letter code is used. The tables \ref{tab:OTbooks} and \ref{tab:NTBooks} in appendix lists the Books in the Bible and the code used for them \cite{USFM}. Each book is further divided into chapters in which the text is split into sentences known as verses. For example, shown below are all the verses (marked at the beginning of sentences) of chapter 23 of the Biblical book of `Psalms' part of the Old Testament (OT).

\begin{quote}{}
\label{qt:Psalms23}
Psalms 23: 

\textsuperscript{1} The Lord is my shepherd; I shall not want. 

\textsuperscript{2} He maketh me to lie down in green pastures: he leadeth me beside the still waters. 

\textsuperscript{3} He restoreth my soul: he leadeth me in the paths of righteousness for his name's sake. 

\textsuperscript{4} Yea, though I walk through the valley of the shadow of death, I will fear no evil: for thou art with me; thy rod and thy staff they comfort me. 

\textsuperscript{5} Thou preparest a table before me in the presence of mine enemies: thou anointest my head with oil; my cup runneth over. 

\textsuperscript{6} Surely goodness and mercy shall follow me all the days of my life: and I will dwell in the house of the Lord for ever.
\end{quote}

To make the dataset easier to use for training automatic models, we perform some post-processing steps and make the data available in the following ways (kept in separate directories): 
\begin{table*}
\centering
\begin{tabular}{llllllll}
\hline
No. & Language & Chapters & Duration & Cleaned- & Duration & Short- & Duration\\
& & & & verses & (cleaned) & verses & (short)\\
\hline
1 & Hindi & 928 & 71.41 & 22751 & 67.62 & 11511 & 22.60\\
2 & Haryanvi & 260 & 27.41 & 7702 & 25.64 & 3037 & 6.29\\
3 & Bilaspuri & 260 & 26.26 & 7163 & 22.67 & 3022 & 6.19\\
4 & Dogri & 260 & 22.28 & 7735 & 20.57 & 4578 & 9.14\\
5 & Gaddi & 260 & 21.81 & 7731 & 20.10 & 4769 & 9.31\\
6 & Bhadrawahi & 260 & 22.16 & 6891 & 18.36 & 4114 & 8.10\\
7 & Kangri & 260 & 22.28 & 7241 & 19.20 & 4368 & 8.65\\
8 & Kulvi & 260 & 25.30 & 7033 & 21.22 & 3319 & 6.76\\
9 & Mandeali & 260 & 25.38 & 6763 & 20.20 & 3353 & 6.83\\
10 & Pahari Mahasui & 260 & 21.33 & 6929 & 17.72 & 4430 & 8.68\\
11 & Kulvi Outer Seraji & 260 & 23.64 & 6995 & 19.62 & 3816 & 7.49\\
\hline
\end{tabular}
\caption{Language wise data size}
\label{tab:datasize table}
\end{table*}

\subsection{Raw Data}

The original audio recordings of the Bible in different languages are stored as one \textit{.mp3} file per chapter of each Bible book. They are named in the pattern \textit{<Book-code\_ chapter-number.mp3>}. The recordings, along with the Bible text, sometimes contain brief introductions to books and chapters as well as recordings of section headings within the chapters. We provide a timestamp file for each chapter’s audio recording file that demarcates the start times of different parts within. The timestamp files have the same name as their corresponding chapter recording file except that they have the \textit{.tsv} file extension. 

The corresponding Bible text is stored in the Universal Scripture Format Markers (\href{https://ubsicap.github.io/usfm/about/index.html}{USFM}) format which is a popular format among Bible translation agencies. For ease of consuming this data, we parse the USFM file and provide the textual content in a \textit{.csv} format. This data is kept in the \textit{raw/text} directory, one file per book, named with the 3-letter book code.

\subsection{Cleaned Data}\label{Cleaned Data}
In order to train ASR models with this data we perform the following pre-processing steps. First, we remove instances of the recordings where we found inconsistencies between the timestamps, text and the audio recording file. Then, we break-up each chapter recording file into separate verse recording files which we store in the \textit{.wav} format. The audio files are named in the pattern \textit{<book-code\_chapter-number\_ verse-number.wav>}, and are referenced in the path field of every dataset file in the \textit{experiments} and the \textit{cleaned} directories. Finally, we remove any verse recording file which is more than 10 seconds in duration, in the \textit{short\_verses}, to support training ASR models that may hit ‘Out of memory’ errors during training otherwise. These are kept in the \textit{cleaned} directory.

\subsection{Experiment Splits}\label{Experiment Splits}
One of our goals in creating this dataset is to use it to build ASR systems for very low resource languages. We want such a model to be trained with minimal data so as to be useful in cases where Bible translations are done orally first. In such a scenario, we assume the recordings of the Bible are available but the text version needs to be produced. Thus, eliciting little manual transcription, we desire to train a reasonable ASR model that can then aid the transcribers with the remaining work while being able to incrementally re-train the model to steadily improve output quality using manually corrected data. 

Another requirement for us to be able to use this model effectively is that it should perform well on a test set that is not similar to the dataset it has been built from in terms of Bible books. The vocabulary and style of language can vary considerably across Bible Books. Thus, it becomes relevant to ensure stable performance of the model when trained on one type of text and then tested on a different type.

We create splits of the cleaned data (\ref{Cleaned Data}) for training and analysing the performance of ASR models. The splits are available in the \textit{experiments} directory.  The file names indicate the experiment and the split category (see Table~\ref{tab:ExpDatasize}). 

The first category of the splits is based on the size of the training data. For this we create  training splits of the sizes 500, 1000 and 2500 verses respectively. Similarly, we create a split with all short verses (those with less than or equal to 10 seconds in length) which contain \textit{\_short} in their filenames. We also create a split with the full data (\textit{\_full}). Each of these are further divided into training and evaluation sets with an 8:2 ratio (e.g. \textit{train\_1000} and \textit{val\_1000}). For these experiments we use a single disjoint test set of 500 verses (\textasciitilde 1 hour) which is specified by the file \textit{test\_common.csv}.

The other category for the splits is based on the difference in writing styles between the Biblical books. For this we create a training (and evaluation) set using only the Gospels (MAT, MRK, LUK and JHN) all of which describe the life of Jesus on earth and then create different test sets from the book of The Acts of the Holy Spirit (ACT- describes events after the resurrection; \textit{test\_acts.csv}), the epistles (ROM, 1CO, 2CO, GAL; \textit{test\_letters.csv}) and the last books in the NT (1JN, 2JN, 3JN, JUD, REV; \textit{test\_lastbooks}). These splits have been created after analyzing the word overlap between the books and identifying groups of most dissimilar books in the NT of these languages. See table \ref{tab:ExpDatasize}, for more information on number of verses in each experiment split.

\section{Experiments}
  
We train two different ASR models and compare results on the training splits defined in section 3.3.

\subsection{Methodology}

\subsubsection{Wav2vec XLS-R}
We use the pretrained wav2vec2 2.0 \cite{wav2vec} based XLS-R model \cite{XLS-R} that was released on HuggingFace \footnote{\url{https://huggingface.co/facebook}} and fine-tune it on our data. XLS-R used almost half a million hours of audio data in 128 languages for self-supervised pre-training and provides pre-trained models with 300 million up to two billion parameters. 
XLS-R learns contextualized speech representations by randomly masking feature vectors before passing them to a transformer network during self-supervised pre-training. For fine-tuning, a single linear layer is added on top of the pre-trained network to train the model on labelled data of the downstream tasks such as speech recognition, speech translation and audio classification. In our use case, which is ASR, we fine-tune the base model separately for each language.  

\textbf{Tokenizer}: After basic pre-processing and cleaning of text data, we create a CTC-tokenizer \cite{ctc} with the following characteristics in its vocabulary: contains all unique letters of the language, the \textit{space} character replaced by “|”, “UNK” token for unknown characters and “PAD” for blank-token as required by the CTC algorithm. 

\textbf{Feature Extractor}: \textit{Wav2Vec2FeatureExtractor} is used with the following configuration, similar to or as required by the base models: Sampling rate 16kHz, feature size 1, padding with 0.0 for shorter inputs, with normalizing and returning attention mask.  

\textbf{Data Collator}: The data collator pads the input dynamically to match sequence size to the longest input in a batch and also treats the input values and labels differently since they are of different modalities. 

\textbf{Evaluation Metric}: Word error rate (WER) is used as the evaluation metric which is the predominant metric in ASR. 

\textbf{The pre-trained checkpoint} of base model is loaded with attention, hidden and feature projection dropout = 0.0, mask time probability=0.05, no layer drops and setting the CTC loss reduction="mean".

And the model’s feature extractor is frozen to avoid further fine-tuning the initial CNN layers that extract acoustically meaningful but contextually independent features from the raw speech signal.

\textbf{Training}: The following values were used alike across all experiments: batches were grouped by length with 8 batches per device, gradient accumulation steps=2, evaluation strategy='steps', gradient checkpointing=True, fp16 training with a learing rate of 3e-4.

The following values were adjusted as per the size of input dataset: training epochs; save, eval and warmup steps.

Long input sequences require a lot of memory. Since XLS-R is based on self-attention, the memory requirements scale quadratically with the input length. In all our experiments we use only verses that are 10 seconds or less in duration except in the 'Full dataset' using which we were not able to train the XLS-R model successfully.

The results shown in table \ref{tab:Results table} are those using the 1billion model\footnote{\url{https://huggingface.co/facebook/wav2vec2-xls-r-1b}} which gave better results than the 300million model\footnote{\url{https://huggingface.co/facebook/wav2vec2-xls-r-300m}} even though for training speed and GPU memory usage the 300million model was found to be better. 

\begin{table*}
\centering
\begin{tabular}{lllllll}
\hline
Language & Dataset & \multicolumn{3}{c}{Wav2vec XLS-R} & \multicolumn{2}{c}{Kaldi} \\	
 & & Eval & Test & Test & Eval & Test \\
 & & & w/o LM & w/ LM & & w/ LM \\

\hline
Hindi & Full data & - & - & - & 2.93 & 3.36\\
 & Short verses & 3.59 & 3.66 & 2.38 & 5.22 & 5.18 \\
 & 2500 & 8.40 & 8.31 & 6.23 & 14.08 & 12.12 \\
 & 1000 & 10.83 & 11.35 & 9.07 & 22.87 & 21.23\\
 & 500 & 12.58 & 14.49 & 12.14 & 37.06 & 36.31\\
\hline
Bilaspuri & Full data &- &- &- & 10.10 & 10.26\\
 & Short verses &14.40 &16.08 & 13.02 &15.31 & 16.16 \\
 & 2500 &16.08 &17.92 &13.9 & 15.47 & 16.21\\
 & 1000 & 18.45 &21.64 &17.96 & 21.64 & 23.58\\
 & 500 & 23.56 & 25.74 & 22.94 & 30.20 & 31.79\\
\hline
Dogri & Full data &- &- &- & 11.89 & 11.91\\
 & Short verses &13.74 &13.59 &11.72 & 14.62 & 14.63\\
 & 2500 &18.68 &17.26 &14.05 & 17.09 & 16.63\\
 & 1000 &18.35 &19.32 &16.56 & 21.26 & 22.95\\
 & 500 &23.81 &25.25 &21.65 & 29.91 & 30.26\\
\hline
Haryanvi & Full data &- &- &- & 16.84 & 18.79\\
 & Short verses & 25.63&23.87 &20.65 & 28.85 & 24.14 \\
 & 2500 &25.67 &24.13 &20.28 & 25.98 & 25.07\\
 & 1000 &33.67 &29.78 &25.43 & 34.48 & 30.45\\
 & 500 &33.15 &30.90 &27.70 & 39.94 & 37.66\\
\hline
\end{tabular}
\caption{WER for the experiments using datasets of different sizes}
\label{tab:Results table}
\end{table*}

\subsubsection{Kaldi Toolkit}
Kaldi \cite{Povey11thekaldi} is a free and open-source speech recognition toolkit licensed under the Apache 2.0 License\footnote{\url{https://www.apache.org/licenses/LICENSE-2.0.html}}. It is primarily written in C++. Important features of Kaldi are: finite state transducer-based framework (using OpenFst toolkit \cite{10.1007/978-3-540-76336-9_3}), extensive linear algebra support with BLAS \cite{blackford2002updated} and LAPACK \cite{lapack99}, extensible design, non-restrictive license, availability of complete recipes for building speech recognition systems and thorough test routines. Adding new feature without modifying the existing modules makes Kaldi's design extensible.

We used the tedlium s5\_r2 recipe\footnote{\url{https://github.com/kaldi-asr/kaldi/blob/master/egs/tedlium/s5_r2}} to train our Hindi acoustic model. Specifically, we used the local/chain/tuning/run\_tdnn\_1g.sh script for preparing and training the TDNN-f\cite{povey18_interspeech} model. We used the Hindi AM as a base model for transfer learning to train the other languages. For transfer learning, we roughly followed the approach used in Kaldi's rm s5 recipe, namely the local/chain/tuning/run\_tdnn\_wsj\_rm\_1c.sh script.
The following steps were involved in building an ASR system using Kaldi.


\begin{table*}
\centering
\begin{tabular}{llllllllll}
\hline
Language & Trainset & \multicolumn{4}{c}{Wav2vec XLS-R} & \multicolumn{4}{c}{Kaldi} \\   
 & & Eval & Acts & Letters & Lastbooks & Eval & Acts & Letters & Lastbooks \\
\hline
Bilaspuri & Gospels &14.58 &22.92 &18.57 &24.80 & 17.02 & 25.66 & 22.78 & 32.1 \\
Dogri & Gospels &12.66 &16.90 &17.38 &26.90 & 15.11 & 22.24 & 21.7 & 30.0 \\
Haryanvi & Gospels &23.24 &34.90 &27.14 &64.37 & 22.65 & 38.04 & 26.9 & 58.1 \\
\hline
\end{tabular}
\caption{WER for the Book-wise experiments}
\label{tab:Gospels Results table}
\end{table*}

\textbf{Data preparation}: Kaldi needs the input audio files and their corresponding transcripts in a particular format for training and testing. For recordings of around 10 seconds duration, we need three files: \textit{wav.scp}, \textit{text}, and \textit{utt2spk}. For recordings of longer duration an additional \textit{segments} file is required. The details of each of these files are: 
\begin{itemize}
    \item \emph{text} file contains the transcriptions of each utterance\\
        Format: <utterance-id> <transcript>
    \item \emph{utt2spk} file says, for each utterance, which speaker (denoted by an ID) spoke it \\
    Format: <utterance-id> <speaker-id> 
    \item \emph{wav.scp} file \\
    Format: <recording-id> <extended-filename> \\
where the "extended-filename" may be an actual filename, or a command that produces a \textit{.wav} file
    \item \emph{Segments} file \\
Format: <utterance-id> <recording-id> <segment-begin> <segment-end> \\
where the segment-begin and segment-end are measured in seconds. 	 
\end{itemize}

\textbf{Lexicon and Language Model}: Typical pronunciation dictionaries map between the orthographical representation of a word and the sequence of phones in its broad phonetic transcription. Typically symbols from systems such as ARPAbet or X-SAMPA are used to represent the phones, but creating a pronunciation dictionary from scratch is a potentially time-consuming and expensive undertaking. Annotators trained in phonetics and the target language must manually write thousands of entries, which can be used to train a grapheme-to-phoneme (G2P) model that can generate automatic pronunciations for the rest of the words in the dataset. Noting that northern Indian languages are quite consistent in their orthography and pronunciation, we decided to use the Unicode\cite{Unicode} characters for Devanagari as our phone set. This allows us to generate the entire lexicon with a simple script, without the potential inconsistencies and errors human annotators and G2P models might introduce. 

The Bible text is used to create a lexicon (or the pronunciation dictionary) by listing all words in the Bible along with its pronunciation, which are its Unicode characters separated by a space character.

\begin{figure}[!ht]
\centering
\includegraphics[scale=.6]{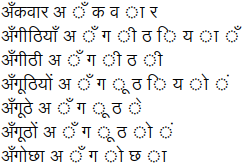}
\caption{Excerpt of the Hindi lexicon used}
\end{figure}

The language model is built using the MIT Language Modelling Toolkit\footnote{\url{https://github.com/mitlm/mitlm}} (MITLM) using the Bible text after removing punctuation and normalizing white space. This toolkit is a set of tools designed for the efficient estimation of statistical n-gram language models. We exclude the evaluation and test set from the language model training and for the low resource languages, we include the Hindi data for training the language model since these are closely related. 

\textbf{Feature extraction}: Kaldi’s feature extraction and waveform-reading module creates standard MFCC (Mel-frequency Cepstral Coefficients) features. Since this module requires the audio recordings in the \textit{.wav} format we convert the \textit{.mp3} files with a 16 KHz sampling frequency.

\textbf{Acoustic modelling}: Once the features are extracted, we train a GMM-HMM based acoustic model to generate alignments (phoneme-to-audio alignments) for the training audio. Using this alignment data, a DNN (Deep Neural Network) based acoustic model is trained. The acoustic model is a factorized time delay neural network (TDNN-f) \cite{povey18_interspeech} with 13 hidden layers, 9882368 parameters, and 3456 outputs. It is a "chain" model, which uses the sequence-level objective function known as lattice-free MMI \cite{povey2016purely}. 


\subsection{Results}

The Word Error Rate (WER) for the ASR models trained on the different data splits are shown in Tables~\ref{tab:Results table}~and~\ref{tab:Gospels Results table}. Due to an 'Out of Memory' error the 'Full data' experiment could not be run using the Wav2vec XLS-R model.

\section{Discussion}
Overall, both the trained ASR models show comparable results for the larger sized datasets (Table~\ref{tab:Results table}). Wav2vec XLS-R outperforms Kaldi on smaller sized datasets. Also, for Hindi, Wav2vec XLS-R shows significant improvements over Kaldi. We attribute both of these observations to the power of pre-training on a large corpus including Hindi data. Also, using a language model, even if not trained on any extraneous text (only the train set) reasonably improves the WER.

We observe lower results in the experiments run for Haryanvi language which we understand is due to the poor data quality, specifically mismatches between the timestamp and audio files.

The experiments on the books with different writing styles show poorer results. Factors that contribute to this include the low word-level overlap and the shift in discourse/content styles between the train and test splits.

\section{Future Work}
We plan to run experiments for the remaining languages in the dataset and continue to release more data in other languages as it becomes available to us from partnering Bible translation agencies. Another direction is training multi-lingual ASR models to overcome limitations of the dataset being single-speaker.



\bibliography{custom}
\bibliographystyle{acl_natbib}

\newpage
\appendix
\label{appendix}
\section{Bible Books and 3-letter Codes}
The tables \ref{tab:OTbooks} and \ref{tab:NTBooks} show the books in the protestant Bible, and their codes used in the dataset.
\begin{table*}[ht]
\centering
\begin{tabular}{llll}
\hline
No. & Book Code & Book Name & Comments\\
\hline
1 & GEN & Genesis & ‘1 Moses’ in some Bibles\\
2 & EXO & Exodus & ‘2 Moses’ in some Bibles\\
3 & LEV & Leviticus & ‘3 Moses’ in some Bibles\\
4 & NUM & Numbers & ‘4 Moses’ in some Bibles\\
5 & DEU & Deuteronomy & ‘5 Moses’ in some Bibles\\
6 & JOS & Joshua &\\
7 & JDG & Judges & \\
8 & RUT & Ruth & \\
9 & 1SA & 1 Samuel  & 1 Kings or Kingdoms in Orthodox Bibles \\
10 & 2SA & 2 Samuel  & 2 Kings or Kingdoms in Orthodox Bibles \\
11 & 1KI & 1 Kings  & 3 Kings or Kingdoms in Orthodox Bibles \\
12 & 2KI & 2 Kings  & 4 Kings or Kingdoms in Orthodox Bibles \\
13 & 1CH & 1 Chronicles  & 1 Paralipomenon in Orthodox Bibles \\
14 & 2CH & 2 Chronicles   & 2 Paralipomenon in Orthodox Bibles \\
15 & EZR & Ezra & This is for Hebrew Ezra (1 Ezra, 1 Esdras)  \\
16 & NEH & Nehemiah  & Sometimes appended to Ezra; called 2 Esdras in the Vulgate  \\
17 & EST & Esther (Hebrew)   & This is for Hebrew Esther; for the longer  
Greek LXX Esther use ESG  \\
18 & JOB & Job  & \\
19 & PSA & Psalms & 150 Psalms in Hebrew  \\
20 & PRO & Proverbs  & 31 Proverbs, but 24 Proverbs in the Ethiopian Bible  \\
21 & ECC & Ecclesiastes  & 3Qoholeth in Catholic Bibles; for Ecclesiasticus use SIR  \\
22 & SNG & Song of Songs   & Song of Solomon, or Canticles of Canticles in  Catholic Bibles  \\
23 & ISA & Isaiah  &  \\
24 & JER & Jeremiah  & The Book of Jeremiah; for the Letter of  
Jeremiah use LJE \\
25 & LAM & Lamentations   &The Lamentations of Jeremiah  \\
26 & EZK & Ezekiel  &  \\
27 & DAN & Daniel (Hebrew)  & This is for Hebrew Daniel; for the longer  
Greek LXX Daniel use DAG \\
28 & HOS & Hosea &\\
29 & JOL & Joel  &  \\
30 & AMO & Amos  &  \\
31 & OBA & Obadiah  &  \\
32 & JON & Jonah  &   \\
33 & MIC & Micah  &  \\
34 & NAM & Nahum  &  \\
35 & HAB & Habakkuk  &  \\
36 & ZEP & Zephaniah  &  \\
37 & HAG & Haggai  &  \\
38 & ZEC & Zechariah  &  \\
39 & MAL & Malachi  &  \\
\hline
\end{tabular}
\caption{Books contained in the Old Testament(OT) of The Bible}
\label{tab:OTbooks}
\end{table*}

\begin{table*}
\centering
\begin{tabular}{llll}
\hline
No. & Book Code & Book Name & Comments\\
\hline
40 & MAT & Matthew  & The Gospel according to Matthew \\
41 & MRK & Mark  &  The Gospel according to Mark \\
42 & LUK & Luke  & The Gospel according to Luke  \\
43 & JHN & John  &  The Gospel according to John \\
44 & ACT & Acts  & The Acts of the Apostles  \\
45 & ROM & Romans  & The Letter of Paul to the Romans  \\
46 & 1CO & 1 Corinthians & The First Letter of Paul to the Corinthians  \\
47 & 2CO & 2 Corinthians  & The Second Letter of Paul to the Corinthians   \\
48 & GAL & Galatians  & The Letter of Paul to the Galatians  \\
49 & EPH & Ephesians  & The Letter of Paul to the Ephesians   \\
50 & PHP & Philippians  & The Letter of Paul to the Philippians   \\
51 & COL & Colossians  & The Letter of Paul to the Colossians   \\
52 & 1TH & 1 Thessalonians   & The First Letter of Paul to the Thessalonians  \\
53 & 2TH & 2 Thessalonians   & The Second Letter of Paul to the  
Thessalonians  \\
54 & 1TI & 1 Timothy   & The First Letter of Paul to Timothy   \\
55 & 2TI & 2 Timothy   & The Second Letter of Paul to Timothy    \\
56 & TIT & Titus  & The Letter of Paul to Titus   \\
57 & PHM & Philemon  & The Letter of Paul to Philemon  \\
58 & HEB & Hebrews  & The Letter to the Hebrews  \\
59 & JAS & James  & The Letter of James   \\
60 & 1PE & 1 Peter   & The First Letter of Peter  \\
61 & 2PE & 2 Peter   & The Second Letter of Peter   \\
62 & 1JN & 1 John    & The First Letter of John   \\
63 & 2JN & 2 John    & The Second Letter of John  \\
64 & 3JN & 3 John    & The Third Letter of John   \\
65 & JUD & Jude   & The Letter of Jude  \\
66 & REV & Revelation  & The Revelation to John; called Apocalypse in  
Catholic Bibles   \\
\hline
\end{tabular}
\caption{Books contained in the New Testament (NT) of The Bible}
\label{tab:NTBooks}
\end{table*}

\section{Sizes of each experiment set}
Table \ref{tab:ExpDatasize} shows the number of verses in each experiment split. The training and validation dataset are split in an 8:2 ratio.
\begin{table*}[ht]
\centering
\begin{tabular}{lcccc}
\hline
Dataset & Hindi & Bilaspuri & Dogri & Haryanvi\\
\hline
all\_verses.csv & 22751 & 7163 & 7735 & 7702\\
short\_verses.csv & 11511 & 3022 & 4578 & 3037\\
\hline
test\_common.csv & 500 & 500 & 500 & 500\\
train\_full.csv & 17800 & 5328 & 5786 & 5757\\
val\_full.csv & 4451 & 1333 & 1447 & 1440\\
train\_short.csv & 8808 & 2017 & 3261 & 2029\\
val\_short.csv & 2203 & 505 & 816 & 508\\
train\_2500.csv & 2000 & 2000 & 2000 & 2000\\
val\_2500.csv & 500 & 500 & 500 & 500\\
train\_1000.csv & 800 & 800 & 800 & 800\\
val\_1000.csv & 200 & 200 & 200 & 200\\
train\_500.csv & 400 & 400 & 400 & 400\\
val\_500.csv & 100 & 100 & 100 & 100\\
\hline
train\_gospels.csv & - & 1360 & 1926 & 1482 \\
val\_gospels.csv & - & 340 & 482 & 371\\
test\_acts.csv & - & 396 & 517 & 334\\
test\_letters.csv & - & 370 & 730 & 356\\
test\_lastbooks.csv & - & 137 & 232 & 132\\
\hline
\end{tabular}
\caption{Number of verses in each dataset file}
\label{tab:ExpDatasize}
\end{table*}



\end{document}